\documentclass[aps,twocolumn,showpacs, preprintnumbers,amsmath,amssymb]{revtex4}

\usepackage{graphicx}
\usepackage{amsmath}

\renewcommand{\rho}{\varrho}
\renewcommand{\phi}{\varphi}
\renewcommand{\vec}{\mathbf}

\begin{document}

\title{Free-energy functional of the Debye-H\"uckel model of simple fluids}

\author{R. Piron$^1$\footnote{Corresponding author\\Electronic address: robin.piron@cea.fr} and T. Blenski$^2$}
\affiliation{$^1$CEA, DAM, DIF, F-91297 Arpajon, France.}
\affiliation{$^2$Laboratoire ``Interactions, Dynamiques et Lasers'', UMR 9222, CEA-CNRS-Universit\'e Paris-Saclay, Centre d'\'Etudes de Saclay, F-91191 Gif-sur-Yvette Cedex, France.}

\date{\today}

\begin{abstract}
The Debye-H\"uckel approximation to the free-energy of a simple fluid is written as a functional of the pair correlation function. This functional can be seen as the Debye-H\"uckel equivalent to the functional derived in the hyper-netted chain framework by Morita and Hiroike, as well as by Lado. It allows one to obtain the Debye-H\"uckel integral equation through a minimization with respect to the pair correlation function, leads to the correct form of the internal energy, and fulfills the virial theorem.
\end{abstract}

\pacs{05.20.Jj}

\maketitle

\section{Introduction}
The Debye-H\"uckel (DH) model originated in the physics of electrolytes \cite{DebyeHuckel23}, that is, fluids of charged particles interacting through a Coulomb potential. It is the linearized version of the classical ``Hartree-like'' mean-field model, which is often called Non-Linear Debye-H\"uckel (NLDH) model or Poisson-Boltzmann model in the context of electrolytes. These two models can be extended to arbitrary interaction potentials, leading respectively to the DH and NLDH models of classical fluids, with the corresponding integral equations. 

Although it is based on very strong assumptions, the DH model gives physical insight into the screening of interaction potentials and the decay of correlation functions, both in the framework of simple and multicomponent fluids. The DH model is valid in the low-coupling limit, that is, in the limit of small interactions compared to the kinetic energy of the particles.
More sophisticated models of classical fluids exist, as for instance the Hyper-Netted Chain (HNC) \cite{Morita58} or Percus-Yevick \cite{PercusYevick58} models, which account for part of the correlations. However, some theoretical studies also proceed by introducing corrections, using the low-coupling DH limit as a starting point (see \cite{Abe59,DeWitt65}).
It was also recently shown that in the DH model, the energy and virial routes are  thermodynamically consistent for any potential \cite{Santos09}. 
For these reasons, the DH model is of permanent theoretical interest.

Moreover, in a number of applications, modified versions of the DH model are used, as for example in the physics of electrolytes \cite{Nordholm84, Penfold90} or in plasma physics \cite{KidderDeWitt61,Vieillefosse81}. The DH model therefore also has some practical interest.

In the case of a long-range attractive potential, the linearization performed in the DH model allows one to circumvent the ``classical catastrophe'' of collapsing particles. To some extent, this explains why this model is so common in plasma physics as well as in the physics of electrolytes.

Apart from the historical derivation as a linearized mean-field theory of a charged-particle mixture \cite{DebyeHuckel23}, the DH model of simple fluids can be obtained from the so-called ``Percus trick'' \cite{Percus64} or from a diagrammatic analysis \cite{Mayer50,Abe59,DeWitt65}. 

In the research on variational models of atoms in plasmas, there is a serious need to account for ion and electron correlations and their impact on the atomic structure and dynamics \cite{Liberman79, Blenski07b, Blenski07a, Piron11, Piron11b, Blenski13, Piron13, Caizergues14, Caizergues16, Starrett12, Starrett12b, Chihara16}. In this framework, it is also useful to have a variational derivation of simple fluid models, using expressions of the free-energy as a functional of the ion-ion pair correlation function. Such a variational expression can be used in a more general theory that also includes the ion electronic structure. 

A free-energy functional is available in the HNC theory \cite{MoritaHiroike60,Lado73}. However, in the well-known DH theory, an expression of the free-energy as a functional of the pair correlation function has not been yet proposed.
The purpose of this paper is to present a brief derivation of such an expression, which can be seen as the DH equivalent to the HNC free-energy functional of Morita and Hiroike \cite{MoritaHiroike60}, also derived by Lado \cite{Lado73}.

\section{Debye-H\"uckel integral equation}
The DH integral equation for a homogeneous simple fluid with the given pair potential $u(r)$ writes (see for example \cite{Percus64}, Eq.~(6.5)):
\begin{equation}\label{eq_DH_int_r}
h(r)=-\beta u(r) - \rho\beta\int d^3r'\left\{h(r')u(|\vec{r}-\vec{r}'|)\right\}
\end{equation}
where $\rho$ is the particle density of the fluid, $\beta$ the inverse temperature, and $h(r)=g(r)-1$, $g(r)$ being the usual pair distribution function.

For some particular pair potentials such as the Coulomb ($1/r$) or Yukawa ($e^{-\alpha r}/r$) potentials, this equation can be recast as a differential equation (Poisson equation in the Coulomb case, Helmholtz equation in the Yukawa case). In a diagrammatic analysis, Eq.~\eqref{eq_DH_int_r} corresponds to the sum of all $u$-bond chain diagrams. 
Eq.~\eqref{eq_DH_int_r} can also be seen as a Ornstein-Zernicke relation with the simplest closure $c(r)=-\beta u(r)$, where $c(r)$ is the direct correlation function.

The Fourier transform $\mathcal{F}_{\vec{k}}$ of a function $\mathcal{F}(\vec{r})$ is defined as:
\begin{equation}
\mathcal{F}_{\vec{k}}=\int d^3r\left\{\mathcal{F}(\vec{r})e^{i\vec{k}.\vec{r}}\right\}
\end{equation}
For a potential that has a Fourier transform, Eq.~\eqref{eq_DH_int_r} can be rewritten in the Fourier space as the algebraic relation:
\begin{equation}\label{eq_DH_int_k}
h_k=-\beta u_k - \rho\beta h_k u_k
\end{equation}
which has the direct formal solution:
\begin{equation}\label{eq_us}
h_{\text{eq},k}=-\beta u_{s,k}\text{ ; }u_{s,k}\equiv\frac{u_k}{1+\rho\beta u_{k}}
\end{equation}
Here, $h_{\text{eq},k}$ denotes the DH equilibrium correlation function and $u_{s,k}$ denotes the screened potential, both in the Fourier space.

For any long-range potential (i.e. Coulombic at infinity), $\lim_{k\rightarrow 0} u_k\sim k^{-2}$, which implies $u_{k=0}=-\rho^{-1}$.
This is the case in any model that can be written as an Ornstein-Zernicke equation with a direct correlation function that includes the interaction potential.

In the following, the pair potential $u(r)$ will be replaced by $u^\xi(r)\equiv\xi u(r)$, $\xi$ being a charge parameter, and the corresponding $\xi$ indices will label $h_\text{eq}^\xi$, $u^\xi$ and $u_s^\xi$.

\section{Expression of the free-energy functional}
We consider a system which is immersed in a rigid, homogeneous, ``neutralizing'' background. Such background do not impact on the equations-of-motion or on the DH integral equation. It however leads to renormalized energies by accounting for the contribution:
\begin{equation}
\frac{E_{bg}}{V}=-\frac{\rho^2}{2}\int d^3r\left\{u^\xi(r)\right\}
\end{equation}
which cancels the divergences of energies in case of a long-range potential. In case of a short-range potential, one can disregard any background by removing this contribution. In the following, we will address the configurational part of the renormalized free-energy and give its expression as a functional of the pair correlation function.

In the particular case of the Coulomb potential: $u^\xi(r)\equiv\xi/r$, it is easy to check that if one minimizes w.r.t. $h(r)$ the following functional $A^\xi\left\{\rho,\beta,h(r)\right\}$:
\begin{align}
\frac{A^\xi}{V}=\frac{2\rho^2}{3\beta} \int d^3r &\left\{h(r)\left(\frac{h(r)}{2}+\frac{\beta\xi}{r}\right.\right.
\nonumber\\
&\left.\left.+\frac{\rho\beta\xi}{2}\int d^3r'\left\{\frac{h(r')}{|\vec{r}-\vec{r}'|}\right\}\right)\right\}
\end{align}
one gets Eq.~\eqref{eq_DH_int_r}. This functional also yields all the well-known equilibrium results for the renormalized configurational internal energy $U_\text{eq}^\xi$, free-energy $A_\text{eq}^\xi$, and pressure $P_\text{eq}^\xi$ (see, for example, \cite{LandauStatisticalPhysics}~\textsection{}78), namely:
\begin{align}
&\frac{U_\text{eq}^\xi}{V}\label{eq_U_DHOCP}
=\frac{\partial}{\partial \beta}\left(\frac{\beta A_\text{eq}^\xi}{V}\right)
=\left.\frac{\partial}{\partial \beta}\left(\frac{\beta A^\xi}{V}\right)\right|_\text{eq}=-\frac{\xi\rho}{2}\sqrt{4\pi\rho\beta\xi}\\
&\frac{A_\text{eq}^\xi}{V}=\frac{2}{3}\frac{U_\text{eq}^\xi}{V}\label{eq_A_DHOCP}\\
&P_\text{eq}^\xi
=\rho^2\frac{\partial}{\partial \rho}\left(\frac{A_\text{eq}^\xi}{\rho V}\right)\label{eq_P_DHOCP}
=\rho^2\frac{\partial}{\partial \rho}\left.\left(\frac{A^\xi}{\rho V}\right)\right|_\text{eq}
=\frac{1}{3}\frac{U_\text{eq}^\xi}{V}
\end{align}
where $|_\text{eq}$ means that the functional is taken at the DH equilibrium, that is, for $h(r)=h_\text{eq}^\xi(r)$ from Eq.~\eqref{eq_us}. We stress that Eqs~\eqref{eq_U_DHOCP},\eqref{eq_A_DHOCP},\eqref{eq_P_DHOCP} hold for a One-component Classical Plasma (OCP) with the contribution of the neutralizing background included.

A slightly more complicated functional can be guessed as well in the Yukawa-OCP case: $u^{\xi}(r)=\xi\exp(-\alpha r)/r$.

The present derivation aims at giving a general form of the free-energy functional for a simple fluid with any pair potential $u(r)$.
As in \cite{Lado73}, we use Debye's charging method and build an expression of the renormalized configurational free-energy functional $A^\xi\left\{\rho,\beta,u(r),h(r)\right\}$ from the equilibrium relation:
\begin{equation}\label{eq_charging_A}
\frac{A_\text{eq}^\xi\left\{\rho,\beta,u(r)\right\}}{V}
=\frac{\rho^2}{2}\int_0^\xi\frac{d\xi'}{\xi'}\int d^3r\left\{h_\text{eq}^{\xi'}(r)u^{\xi'}(r)\right\}
\end{equation}
In the case of the exact equilibrium, Eq.~\eqref{eq_charging_A} is an exact relation. We consider the case of $|_\text{eq}$ denoting the DH equilibrium, with $h(r)=h_\text{eq}^\xi(r)$ from Eq.~\eqref{eq_us}, and require Eq.~\eqref{eq_charging_A} to hold.

Let us assume that the free-energy functional $A^\xi\left\{\rho,\beta,u(r),h(r)\right\}$ can be written in the form that follows:
\begin{equation}
\frac{A^\xi}{V}=\int\frac{d^3k}{(2\pi)^3}\left\{f_k^\xi h_k\left(\frac{h_k}{2} + \beta u_k^\xi + \frac{\rho\beta}{2}h_k u_k^\xi\right)\right\}
\end{equation}
where $f_k^\xi\equiv f(\rho,\beta,\xi,k,u_k^\xi)$ is independent of $h_k$. Minimization of $A^\xi$ w.r.t. $h(r)$ then lead to Eq.~\eqref{eq_DH_int_k}, i.e. to the DH equation.

Differentiating w.r.t. $\xi$, we obtain:
\begin{align}
\frac{\xi}{V}\frac{\partial A_\text{eq}^\xi}{\partial \xi}
=& \left.\frac{\xi}{V}\frac{\partial A^\xi}{\partial \xi}\right|_\text{eq}\\
=&\int\frac{d^3k}{(2\pi)^3}\left\{\xi\frac{\partial f_k^\xi}{\partial \xi} h_k\left(\frac{h_k}{2} + \beta u_k^\xi + \frac{\rho\beta}{2}h_k u_k^\xi\right)\right.\nonumber\\
&+\left.\left.f_k^\xi h_k\left(\beta \xi\frac{\partial u_k^\xi}{\partial \xi} + \frac{\rho\beta}{2}h_k \xi\frac{\partial u_k^\xi}{\partial \xi}\right)\right\}\right|_\text{eq}
\end{align}
Using $u_k^\xi=\xi u_k$, and substituting the formal equilibrium solution of Eq.~\eqref{eq_us}, we get:
\begin{align}
\frac{\xi}{V}\frac{\partial A_\text{eq}^\xi}{\partial \xi}
=&\frac{-\beta^2}{2}\int\frac{d^3k}{(2\pi)^3}\left\{
\frac{\partial}{\partial \xi} \left(\xi f_k^\xi\right) u_{s,k}^\xi u_k^\xi + f_k^\xi u_{s,k}^{\xi\,2}
\right\}
\end{align}

On the other hand, from Eqs.~\eqref{eq_charging_A} and \eqref{eq_us} we can write:
\begin{equation}\label{eq_charging_A_bis}
\frac{\xi}{V}\frac{\partial A_\text{eq}^\xi}{\partial \xi}=\frac{-\beta\rho^2}{2}\int \frac{d^3k}{(2\pi)^3}\left\{u_{s,k}^{\xi} u^{\xi}_k\right\}
\end{equation}

Requiring Eq.~\eqref{eq_charging_A} (or equivalently Eq.~\eqref{eq_charging_A_bis}) to hold, results in the equation:
\begin{align}
\frac{\partial}{\partial \xi}\left(\xi f_k^\xi\right) + \frac{\xi f_k^\xi}{\xi+\rho\beta u_k \xi^2}=\frac{\rho^2}{\beta}
\end{align}
A solution to this differential equation for $\xi f_k^\xi$ can be looked after in the form of a series. This leads to the following expression for $f_k^\xi$:
\begin{equation}\label{eq_sol_fkxi}
f_k^\xi=\frac{\rho^2}{\beta}F(\rho\beta u_k^\xi)
\end{equation}
where the function $F$ is defined as the series:
\begin{equation}
F(x)=1-\sum_{n=0}^\infty\left\{\frac{(-x)^n}{(n+1)(n+2)}\right\}
\end{equation}
The latter can be identified with the function:
\begin{equation}
F(x)=\left(1+\frac{1}{x}\right)\left(1-\frac{\ln(1+x)}{x}\right)
\end{equation}
As can be seen in Eq.~\eqref{eq_sol_fkxi}, the only dependence of $f_k^\xi$ on $k$ and $\xi$  is via the potential $u_k^\xi$.

Finally, the DH free-energy functional reads:
\begin{align}\label{eq_explicit_functional}
\frac{A^\xi}{V}=\frac{\rho^2}{\beta}\int\frac{d^3k}{(2\pi)^3}\left\{
\left(1+\frac{1}{\rho\beta u_k^\xi}\right)\left(1-\frac{\ln(1+\rho\beta u_k^\xi)}{\rho\beta u_k^\xi}\right)
\right.\nonumber\\\left.
h_k\left(\frac{h_k}{2}+\beta u_k^\xi+\frac{\rho\beta}{2}h_ku_k^\xi\right)\right\}
\end{align}
which is the main result of the present paper.

It is worth to note the appearance of the logarithm function, both in the DH and in the HNC free-energy functionals \cite{MoritaHiroike60,Lado73}, which is related to the fact that the Ornstein-Zernicke relation is explicitly fulfilled in both approaches.

\section{Internal energy, pressure and virial theorem}
We first consider the renormalized configurational internal energy, as defined by the thermodynamics:
\begin{equation}
\frac{U^\xi_\text{eq}}{V}=\frac{\partial}{\partial \beta}\left(\frac{\beta A^\xi_\text{eq}}{V}\right)=\left.\frac{\partial}{\partial \beta}\left(\frac{\beta A^\xi}{V}\right)\right|_\text{eq}
\end{equation}
As can be seen in Eq.~\eqref{eq_explicit_functional}, $\beta A^\xi/V$ is a functional of $h(r)$, $\rho$, and of the product $\beta \xi u(r)$. As a consequence, we can write:
\begin{equation}
\frac{\partial}{\partial \beta}\left(\frac{\beta A^\xi}{V}\right)=\xi\frac{\partial}{\partial \xi}\left(\frac{A^\xi}{V}\right)
\end{equation}
We then have, in virtue of Eq.~\eqref{eq_charging_A}:
\begin{equation}\label{eq_U_eq}
\frac{U^\xi_\text{eq}}{V}=\left.\xi\frac{\partial}{\partial \xi}\left(\frac{A^\xi}{V}\right)\right|_\text{eq}
=\int d^3r\left\{h_\text{eq}^{\xi}(r)u^{\xi}(r)\right\}
\end{equation}
which corresponds to the exact expression of the internal energy, with $h_\text{eq}^{\xi}(r)$ taken to be the DH approximation to the equilibrium correlation function.

The result Eq.~\eqref{eq_U_eq} can also be obtained in a more pedestrian way, by making the explicit differentiation of the free-energy expression Eq.~\eqref{eq_explicit_functional} and then substituting $h_{\text{eq},k}^\xi$. We do not reproduce this calculation here.

The virial configurational pressure as a functional of $\rho$,$u(r)$,$h(r)$ is:
\begin{equation}\label{eq_def_virialP}
P_v^\xi=-\frac{1}{3}\frac{\rho^2}{2}\int d^3r\left\{h(r) \vec{r}.\vec{\nabla}_\vec{r}u^\xi(r)\right\}
\end{equation}
For a potential that decays as $1/r$ or faster, using the Fourier representation and integrating by part, we can write from Eq.~\eqref{eq_def_virialP}: 
\begin{equation}
P_v^\xi=\int d^3r\left\{h(r)u^{\xi}(r)\right\}+\frac{1}{3}\frac{\rho^2}{2}\int \frac{d^3k}{(2\pi)^3}\left\{h_k \vec{k}.\vec{\nabla}_\vec{k}u^\xi_k\right\}
\end{equation}
Using Eq.~\eqref{eq_us}, the equilibrium virial pressure becomes:
\begin{equation}
P_{v,\text{eq}}^\xi=\frac{U^\xi_\text{eq}}{V}
-\frac{\rho^2}{2}\int_0^\infty \frac{dk}{(2\pi)^3}\left\{\frac{4\pi}{3}k^3\frac{\beta u_{k}^\xi}{1+\rho\beta u_{k}^\xi} \frac{\partial u_k^\xi}{\partial k}\right\}
\end{equation}
Finally, we obtain:
\begin{equation}
P_{v,\text{eq}}^\xi=
\frac{U^\xi_\text{eq}}{V}+\frac{1}{2\beta}\int \frac{d^3k}{(2\pi)^3}\left\{\rho\beta u^\xi_k-\ln\left(1+\rho\beta u^\xi_k\right)\right\}
\end{equation}
Then, differentiating w.r.t. $\xi$, we get:
\begin{equation}\label{eq_diff_Pvir}
\xi\frac{\partial P_{v,\text{eq}}^\xi}{\partial \xi}
=\frac{-\rho^2\beta}{2}\int \frac{d^3k}{(2\pi)^3}\left\{\left(u^{\xi}_{s,k}\right)^2\right\}
\end{equation}

We now consider the equilibrium pressure, as it is defined by the thermodynamics:
\begin{equation}
P_\text{eq}^\xi=\rho^2\frac{\partial}{\partial \rho}\left(\frac{A_\text{eq}^\xi}{\rho V}\right)
\end{equation}
Using Eq.~\eqref{eq_charging_A}, we obtain:
\begin{equation}
P_\text{eq}^\xi
=\rho^2\frac{\partial}{\partial \rho} \left( \frac{-\beta\rho}{2}
\int_0^\xi \frac{d\xi'}{\xi'}
\int \frac{d^3k}{(2\pi)^3}\left\{u_{s,k}^{\xi'} u^{\xi'}_k\right\} \right)
\end{equation}
Performing the differentiation w.r.t. $\rho$, we end up with:
\begin{equation}
P_\text{eq}^\xi
=\frac{-\rho^2\beta}{2}\int_0^\xi \frac{d\xi'}{\xi'}\int \frac{d^3k}{(2\pi)^3}\left\{\left(u^{\xi}_{s,k}\right)^2\right\}
\end{equation}
Finally, using Eq.~\eqref{eq_diff_Pvir}, we have:
\begin{equation}\label{eq_virial}
P_\text{eq}^\xi
=\int_0^\xi \frac{d\xi'}{\xi'}\left\{\xi'\frac{\partial P_{v,\text{eq}}^{\xi'}}{\partial \xi'}\right\}
=P_{v,\text{eq}}^{\xi}
\end{equation}
Thus, using the exact Eq.~\eqref{eq_charging_A} in order to define the free-energy in the approximate DH case ensures that the virial theorem is fulfilled.

Again, the result Eq.~\eqref{eq_virial} can also be obtained in a more pedestrian way, from the explicit differentiation of the free-energy expression of Eq.~\eqref{eq_explicit_functional} and the substitution of $h_{\text{eq},k}^\xi$. We do not reproduce this calculation here either.

\section{Conclusion}

Our Eq.~\eqref{eq_explicit_functional} is an explicit expression of the renormalized configurational free-energy functional in the Debye-H\"uckel approximation. This expression allows one to obtain the Debye-H\"uckel equation from a minimization procedure w.r.t. the pair correlation function. In the DH case, as in the HNC case of \cite{MoritaHiroike60,Lado73}, requiring the exact charging relation to hold in the approximate model allows one to define a free-energy functional that yields the correct expression for the internal energy and fulfills the virial theorem.
Work is now in progress in order to generalize the present formalism to multicomponent fluids, as it was done in the HNC case \cite{Lado73b,Enciso87}.

\bibliographystyle{unsrt}
\bibliography{draft}

\begin{thebibliography}{10}

\bibitem{DebyeHuckel23}
P.~Debye and E.~H\"{u}ckel.
\newblock {Zur Theorie des Elektrolyte. I. Gefrierpunktserniedrigung und
  verwandte Erscheinungen}.
\newblock {\em Phys. Z.}, 24:185--206, 1923.

\bibitem{Morita58}
T.~Morita.
\newblock {Theory of Classical Fluids: Hyper-Netted Chain Approximation, I}.
\newblock {\em Prog. Theor. Phys.}, 20:920--938, 1958.

\bibitem{PercusYevick58}
J.~K. Percus and G.~J. Yevick.
\newblock {Analysis of Classical Statistical Mechanics by Means of Collective
  Coordinates}.
\newblock {\em Phys. Rev.}, 110:1--13, 1958.

\bibitem{Abe59}
R.~Abe.
\newblock {Giant Cluster Expansion Theory and Its Application to High
  Temperature Plasma}.
\newblock {\em Prog. Theor. Phys.}, 22:213--226, 1959.

\bibitem{DeWitt65}
H.~E. DeWitt.
\newblock {Classical Theory of the Pair Distribution Function of Plasmas}.
\newblock {\em Phys. Rev.}, 140:A466--A470, 1965.

\bibitem{Santos09}
A.~Santos, R.~Fantoni, and A.~Giacometti.
\newblock {Thermodynamic consistency of energy and virial routes: An exact
  proof within the linearized Debye-H\"uckel theory}.
\newblock {\em J. Chem. Phys.}, 131:181105, 2009.

\bibitem{Nordholm84}
S.~Nordholm.
\newblock {Simple Analysis of the thermodynamic properties of the one-component
  plasma}.
\newblock {\em Chem. Phys. Lett.}, 105:302--306, 1984.

\bibitem{Penfold90}
R.~Penfold, S.~Nordholm, B.~J\"{o}nsson, and C.~E. Woodward.
\newblock {A simple analysis of ion-ion correlation in polyelectrolyte
  solutions}.
\newblock {\em J. Chem. Phys.}, 92:1915--1922, 1990.

\bibitem{KidderDeWitt61}
R.~E. Kidder and H.~E. DeWitt.
\newblock {Application of a modified Debye-H\"{u}ckel theory to fully ionized
  gases}.
\newblock {\em J. Nucl. Energy, Part C: Plasma Physics}, 2:218--223, 1961.

\bibitem{Vieillefosse81}
P.~Vieillefosse.
\newblock {Improved Debye H\"{u}ckel theory for one- and multicomponent
  plasmas}.
\newblock {\em J. Physique}, 42:723--733, 1981.

\bibitem{Percus64}
J.~K. Percus.
\newblock {\em {The equilibrium theory of classical fluids}}, chapter The pair
  distribution function in classical statistical mechanics, pages II--33 --
  II--170.
\newblock W. A. Benjamin, Inc., New York, 1964.

\bibitem{Mayer50}
J.~E. Mayer.
\newblock {The Theory of Ionic Solutions}.
\newblock {\em J. Chem. Phys.}, 18:1426--1436, 1950.

\bibitem{Liberman79}
D.~A. Liberman.
\newblock {Self-consistent field model for condensed matter}.
\newblock {\em Phys. Rev. B}, 20(12):4981--4989, 1979.

\bibitem{Blenski07b}
T.~Blenski and B.~Cichocki.
\newblock {Variational theory of average-atom and superconfigurations in
  quantum plasmas}.
\newblock {\em Phys. Rev. E}, 75:056402, 2007.

\bibitem{Blenski07a}
T.~Blenski and B.~Cichocki.
\newblock {Variational approach to the average-atom-in-jellium and
  superconfigurations-in-jellium models with all electrons treated
  quantum-mechanically}.
\newblock {\em High Energy Density Physics}, 3:34--47, 2007.

\bibitem{Piron11}
R.~Piron and T.~Blenski.
\newblock {Variational-average-atom-in-quantum-plasmas (VAAQP) code and virial
  theorem: Equation-of-state and shock-Hugoniot calculations for warm dense Al,
  Fe, Cu, and Pb}.
\newblock {\em Phys. Rev. E}, 83:026403, 2011.

\bibitem{Piron11b}
R.~Piron and T.~Blenski.
\newblock {Variational average-atom in quantum plasmas (VAAQP) -- Recent
  progress, virial theorem and applications to the equation-of-state of warm
  dense Be}.
\newblock {\em High Energy Density Physics}, 7:346--352, 2011.

\bibitem{Blenski13}
T.~Blenski, R.~Piron, C.~Caizergues, and B.~Cichocki.
\newblock {Models of atoms in plasmas based on common formalism for bound and
  free electrons}.
\newblock {\em High Energy Density Physics}, 9:687--695, 2013.

\bibitem{Piron13}
R.~Piron and T.~Blenski.
\newblock {Variational Average-Atom in Quantum Plasmas (VAAQP) -- Application
  to radiative properties}.
\newblock {\em High Energy Density Physics}, 9:702 -- 710, 2013.

\bibitem{Caizergues14}
C.~Caizergues, T.~Blenski, and R.~Piron.
\newblock {Linear response of a variational average atom in plasma:
  Semi-classical model}.
\newblock {\em High Energy Density Physics}, 12:12 -- 20, 2014.

\bibitem{Caizergues16}
C.~Caizergues, T.~Blenski, and R.~Piron.
\newblock {Dynamic linear response of atoms in plasmas and photo-absorption
  cross-section in the dipole approximation}.
\newblock {\em High Energy Density Physics}, 18:7 -- 13, 2016.

\bibitem{Starrett12}
C.~E. Starrett and D.~Saumon.
\newblock {Fully variational average atom model with ion-ion correlations}.
\newblock {\em Phys. Rev. E}, 85:026403, 2012.

\bibitem{Starrett12b}
C.E. Starrett and D.~Saumon.
\newblock {A variational average atom approach to closing the quantum
  Ornstein--Zernike relations}.
\newblock {\em High Energy Density Physics}, 8:101--104, 2012.

\bibitem{Chihara16}
J.~Chihara.
\newblock {Average Atom Model based on Quantum Hyper-Netted Chain Method}.
\newblock {\em High Energy Density Physics}, 19:38--47, 2016.

\bibitem{MoritaHiroike60}
T.~Morita and K.~Hiroike.
\newblock {A New Approach to the Theory of Classical Fluids. I}.
\newblock {\em Prog. Theor. Phys.}, 23:1003--1027, 1960.

\bibitem{Lado73}
F.~Lado.
\newblock {Perturbation Correction for the Free Energy and Structure of Simple
  Fluids}.
\newblock {\em Phys. Rev. A}, 8:2548--2552, 1973.

\bibitem{LandauStatisticalPhysics}
L.~D. Landau and E.~M. Lifshitz.
\newblock {\em {Statistical Physics}}.
\newblock Pergamon Press, 1959.

\bibitem{Lado73b}
F.~Lado.
\newblock {Perturbation Correction for the Free Energy and Structure of Simple
  Fluid Mixtures}.
\newblock {\em J. Chem. Phys.}, 59:4830--4835, 1973.

\bibitem{Enciso87}
E.~Enciso, F.~Lado, M.~Lombardero, J.~L.~F. Abascal, and S.~Lago.
\newblock {Extension of the optimized RHNC equation to multicomponent liquids}.
\newblock {\em J. Chem. Phys.}, 87:2249--2255, 1987.

\end{thebibliography}

\end{document}